\documentclass[10pt,aps,prb,twocolumn,noshowpacs]{revtex4-1}
\usepackage{amsmath,graphicx,amsbsy,amssymb,amsmath,epsfig,latexsym,wasysym,pifont,mathrsfs,natbib}
\usepackage{float} \newfloat{widefig}{thp}{lop} \usepackage{comment}

\usepackage{array, makecell} \usepackage{verbatim} \usepackage{datetime}
\usepackage{multirow,array} \usepackage{hyperref} \usepackage{color} \usepackage{setspace}
\usepackage{subcaption}
\usepackage{graphicx}
\usepackage[export]{adjustbox}

\hypersetup{colorlinks,
  linkcolor=blue,%
  citecolor=blue,%
  urlcolor=blue}

\begin{document}
\UseRawInputEncoding
\title{\textbf{In-plane ordering and tunable magnetism in Cr-based MXenes}}

\author{Himangshu Sekhar Sarmah}
\email[]{shimangshu@iitg.ac.in}
\affiliation{Department of Physics, Indian Institute of Technology
  Guwahati, Guwahati-781039, Assam, India.}    
\author{Subhradip Ghosh}
\email{subhra@iitg.ac.in} \affiliation{Department of Physics,
  Indian Institute of Technology Guwahati, Guwahati-781039, Assam,
  India.} 

\begin{abstract}
MXene, the two-dimensional derivatives of MAX compounds, due to their structural and compositional flexibility, is an ideal family of compounds to study a number of structure-property relations. In this work, we have investigated the tunability of magnetic properties in Cr-based MXenes that have an in-plane ordering arising out of alloying Cr with another non-magnetic transition metal atom. Using Density Functional Theory based calculations we have explored the effects of composition and surface functionalisations on the electronic and magnetic properties of these in-plane ordered MXenes known as i-MXenes. We found that the electronic and magnetic ground states  are quite sensitive to the structure and composition. This provides enough tunability in these compounds so that they can be used for practical applications. Our calculated results of magnetic transition temperatures and magnetic anisotropy energies are comparable to many a established two-dimensional magnets. These put together widen the prospect of these i-MXenes for multiple usage as magnetic devices  making them attractive for further investigation. 
\end{abstract} 

\maketitle

\section{Introduction\label{intro}}
Two-dimensional (2D) transition metal carbides or nitrides, known as MXene, with general formula M\textsubscript{n+1}X\textsubscript{n}T\textsubscript{n+1}(M: group III-IV transition metal, X: C or N, T: functional groups such as -F, -O,-OH, $n$: integer number) \citep{naguib2011two} derived from their precursor MAX phases M\textsubscript{n+1}AX\textsubscript{n}(A: main group element) by selective etching of the A-atoms, have drawn considerable attention due to the promises shown in multitude of applications \cite{lukatskaya2017ultra,lukatskaya2013cation,fan2018modified,ibragimova2021surface,scheibe2019study,sarikurt2018influence,naguib2013new,ahmed2016h,du2017environmental,shuai2023recent,bai2021recent}. The flexibility in the composition and the presence of functional groups T passivating the bonds dangling on their surfaces \cite{hope2016nmr} allow tunability of the target properties in MXene compounds.

In addition to conventional MXenes, the possibility of alloying on the M,X, or T sites opens up new avenues to explore structure-property relations over a large database of compounds, enhancing their importances from the application and scientific viewpoint. That such a route can indeed give rise to interesting physical phenomena and superior functional behaviour has been observed recently \cite{ourpaper1,ourpaper2,ourpaper3} where alloying led to chemically disordered as well as out-of-plane ordered structures (o-MXene) with reduced symmetry. In-plane chemically ordered phases (i-MAX phase) are recently observed in MAX compounds where alloying with a specific composition is done at the M site. The i-MAX compounds with chemical formula (M\textsubscript{2/3}M'\textsubscript{1/3})\textsubscript{2}AX \cite{tao2017two,dahlqvist2017prediction}can be stabilised in a monoclinic structure (space group C2/c) \cite{tao2017two} if a difference between the atomic radii of M and M$^{\prime}$ is at least 0.2 \AA (M$^{\prime}>$M) \cite{ahmed2020}. Till date, more than 30 i-MAX compounds have been synthesised \cite{tao2017two,dahlqvist2017prediction,meshkian2018w,lu2017theoretical,chen2018theoretical,ahmed2020}. Out of them a few i-MXenes have been exfoliated \cite{meshkian2018w,zhan2019,ahmed2020}. These i-MXenes have exhibited promising potentials as energy storage devices and catalysts in hydrogen evolution reactions \cite{ahmed2020}, opening up the possibility to synthesise and explore more of them in a variety of applications. 

Magnetism is one possible area in which  i-MXenes can be quite useful. The tunability with regard to the transition metal atom in these compounds offers a lot of scope to obtain various magnetic structures exhibiting multiple functional properties associated with magnetism. This is particularly important in the backdrop of   
two-dimensional (2D) magnetic materials garnering immense interest in recent years \cite{huang2017layer,wang2021ferromagnetism,gong2017discovery,lee2016ising}. Though no i-MXene with magnetic transition metal atom as constituents is yet synthesised, first-principles Density Functional Theory \cite{dft} based calculations on more than 300 i-Mxenes \cite{gao2020magnetic} proposed several magnetic i-MXenes exhibiting superior functional properties. A multiferroic i-MXene (Ta\textsubscript{2/3}Fe\textsubscript{1/3})\textsubscript{2}C \cite{zhao2021multiferroic} was also proposed from the results of DFT calculations. In these two works, there are two aspects which are not addressed: first, in both works the magnetic constituent was M$^{\prime}$, proportionally half of the non-magnetic transition metal constituent M, and second, the effects of different functional groups passivating the surfaces of i-MXenes were not considered. Reference \onlinecite{gao2020magnetic}, the most comprehensive work on the i-MXenes, did not consider the role of the functional groups in spite of the well-known facts that the presence of them are inevitable, given the status of experimental techniques to exfoliate i-MXenes from the corresponding i-MAX phases and that they can influence the electronic properties significantly. 

To address these issues, in this work, we have employed DFT based techniques to investigate the structural, electronic and magnetic properties of Cr-based i-MXenes, specifically (Cr\textsubscript{2/3}Y\textsubscript{1/3})\textsubscript{2}C, (Cr\textsubscript{2/3}Zr\textsubscript{1/3})\textsubscript{2}C, and (Cr\textsubscript{2/3}Sc\textsubscript{1/3})\textsubscript{2}C.The reason behind choice of these three compounds is (1) the corresponding i-MAX compounds (Cr\textsubscript{2/3}Zr\textsubscript{1/3})\textsubscript{2}AlC, (Cr\textsubscript{2/3}Sc\textsubscript{1/3})\textsubscript{2}AlC and (Cr\textsubscript{2/3}Y\textsubscript{1/3})\textsubscript{2}AlC have been synthesised \cite{chen2018theoretical,lu2017theoretical} suggesting possibility of exfoliation to i-MXene phases, (2) the magnetic element Cr has a larger proportion than the transition metal element M$^{\prime}$ and (3) DFT calculations of the i-MAX phases of these compounds reveal that the magnetic ground states change across the compounds inspite of the strong magnetic element Cr being present in same proportion. In our work, we investigate, apart from the electronic and magnetic ground states, the finite temperature properties and the magnetic anisotropy energies for pristine, O- and F- functionalised i-MXenes. Our results provide important insights into understanding of the magnetism in this new class of compounds and their operational usefulness. 

\section{Computational Details}
\begin{figure*}
    \includegraphics[height=7cm, width=17.00 cm]{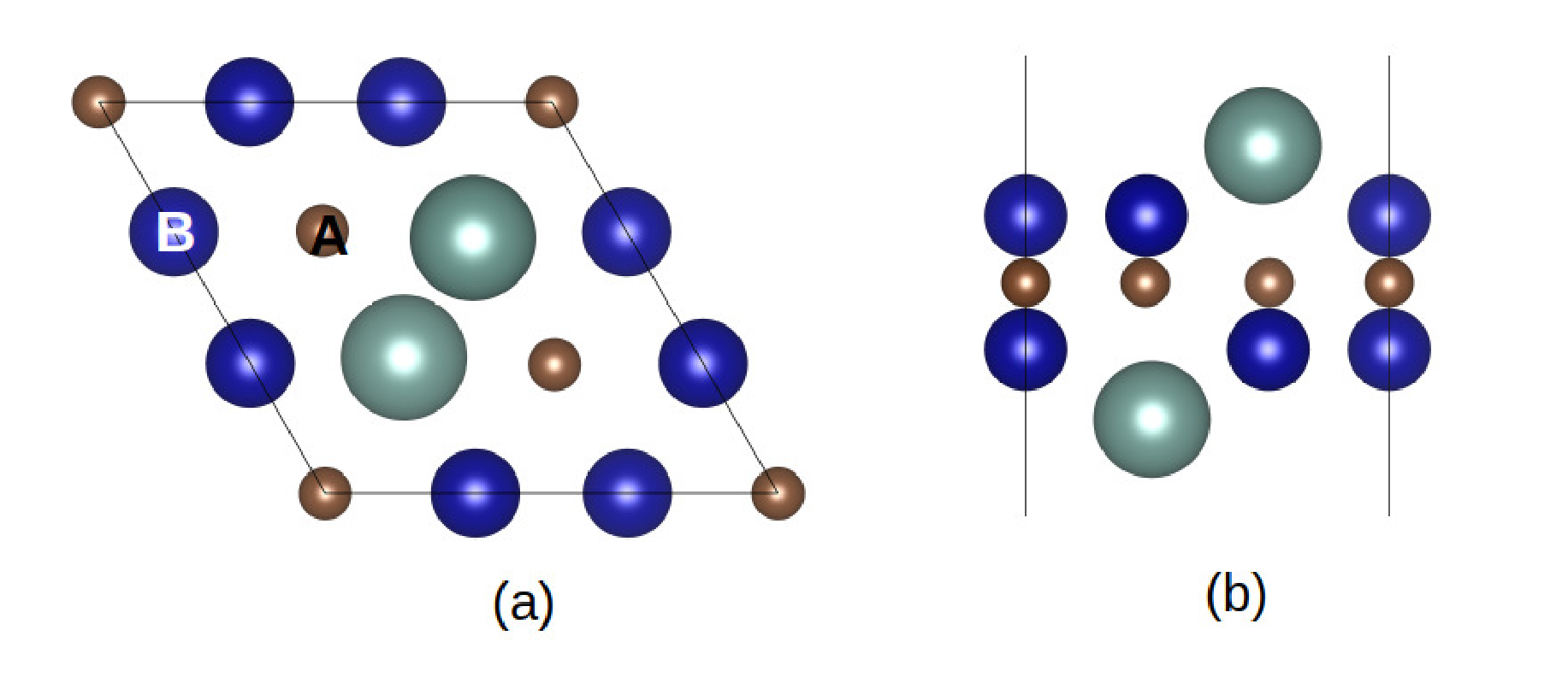}
     \caption{Conventional cell of (Cr$_{2/3}$M$_{1/3})_{2}$C i-MXene. Figures (a) and (b) show top and side views, respectively. Blue, brown and green balls stand for Cr,C and M$^{\prime}$ atoms, respectively.}
    \label{Fig:1}
\end{figure*}
All calculations in this work are performed using first-principles electronic structure methods based on Density Functional Theory (DFT)\cite{dft}, as implemented in the Vienna ab initio simulation (VASP) \citep{kresse1996efficient} package. Projector augmented wave (PAW)\citep{kresse1999ultrasoft} pseudopotentials and a plane wave basis set upto 600 eV are used. The exchange-correlation part of the Hamiltonian is approximated using the generalized gradient approximation (GGA)  parametetrized by Perdew-Burke-Ernzerhof (PBE)\citep{perdew1996generalized}. Van Der Waals interactions are included using DFT-D3 method \citep{grimme2010consistent}. The convergence criteria for the energy and the force are set as $10^{-6}$ eV and 0.01 eV/\AA, respectively.… A $11\times 11\times1$ Monkhorst-Pack \citep{monkhorst1976special} $k$-point mesh is used for geometry optimization. A larger $k$-mesh of $37\times 37\times1$ is used for electronic structure calculations. Spin-orbit coupling (SOC) is included for calculation of the magnetic anisotropy energy (MAE). To prevent interactions between adjacent layers, a 20 \AA  vacuum is  considered.

Two important quantities associated with the formation of the i-MXenes, the exfoliation energy $E_{exf}$ \cite{dolz2021exfoliation}and the formation energy $E_{form}$ are calculated the following way
\begin{align*}
\resizebox{\hsize}{!}{$E_{exf} =\frac{-[E_{total}(i-MAX)-2E_{total}(i-MXene)-2E_{total}(Al)]}{4S} $}.
\end{align*} 
\begin{align*}
\resizebox{\hsize}{!}{$E_{form} =\frac{(E[(Cr_{2/3}M^{\prime}_{1/3})_{2}CT_{2} ]-E[(Cr_{2/3}M^{\prime}_{1/3})C] -6E(T_{g})}{6} $}.
\end{align*}

 $E[(Cr_{2/3}M^{\prime}_{1/3})_{2}CT_{2} ],E[(Cr_{2/3}M^{\prime}_{1/3})C], E(T_{g})$ are the total energies of i-MXenes functionalized by $T$, pristine i-MXene and that of functional group $T$ in gas phase, respectively.. The factor of $6$ is due to the presence of 6 $X$ atoms in the unit cell. $E_{total}(i-MAX), E_{total}(i-MXene),E_{total}(Al)$ are the total energies of the precursor i-MAX, the exfoliated pristine i-MXene and Aluminium in its bulk phase, respectively. $S$ is the surface area of the exfoliated i-MXene.

In order to understand the ground state magnetic structures and subsequently calculate finite temperature properties, one needs computation of the magnetic pair exchange parameters. We have calculated the exchange parameters using full-potential linear muffin-tin orbital (FP-LMTO) method implemented in Relativistic Spin Polarised Toolkit (RSPt)\cite{wills2010full}. In this approach, the spin part of the Hamiltonian is mapped to a Heisenberg model:
\begin{eqnarray}
H = -\sum_{\mu,\nu}\sum_{i,j}
J^{\mu\nu}_{ij}
\mathbf{e}^{\mu}_{i}
.\mathbf{e}^{\nu}_{j}
\end{eqnarray}
$\mu$, $\nu$ represent different sub lattices; \emph{i}, \emph{j} represent atomic positions; and $\mathbf{e}^{\mu}_{i}$ denotes the unit vector along the direction of magnetic moments at site \emph{i} belonging to sub-lattice $\mu$. The $J^{\mu \nu}_{ij}$s are calculated
from the energy differences due to infinitesimally small orientations of a pair of spins using the Magnetic Force Theorem \cite{liechtenstein1987local}. In order to calculate the energy
differences by the RSPt code, a full-potential spin polarized scaler relativistic Hamiltonian with angular momentum cut off $l_{max} = 8$
is used along with a converged $k$ mesh for Brillouin zone integrations. The energy convergence criterion is set to 10$^{-8}$ eV for the self-consistency
cycles. The Matsubara frequency is set to be 1500 after carefully checking its convergence. Using these exchange parameters the magnetic transition temperatures are estimated by the Classical Monte Carlo simulation (MCS) method as implemented in the Uppsala Atomic Spin Dynamics (UppASD) code \cite{eriksson2017atomistic}. In here, calculations are performed for three ensembles in supercell size of $45 \times 45 \times1$. Periodic boundary condition is invoked in each case.  $10^{5}$ Monte Carlo steps are performed at each  temperature to achieve convergence in energies .
\section{Results and discussion}
\subsection{Structural models, and energetics}
\begin{figure*}
    \includegraphics[height=9cm, width=17.00 cm]{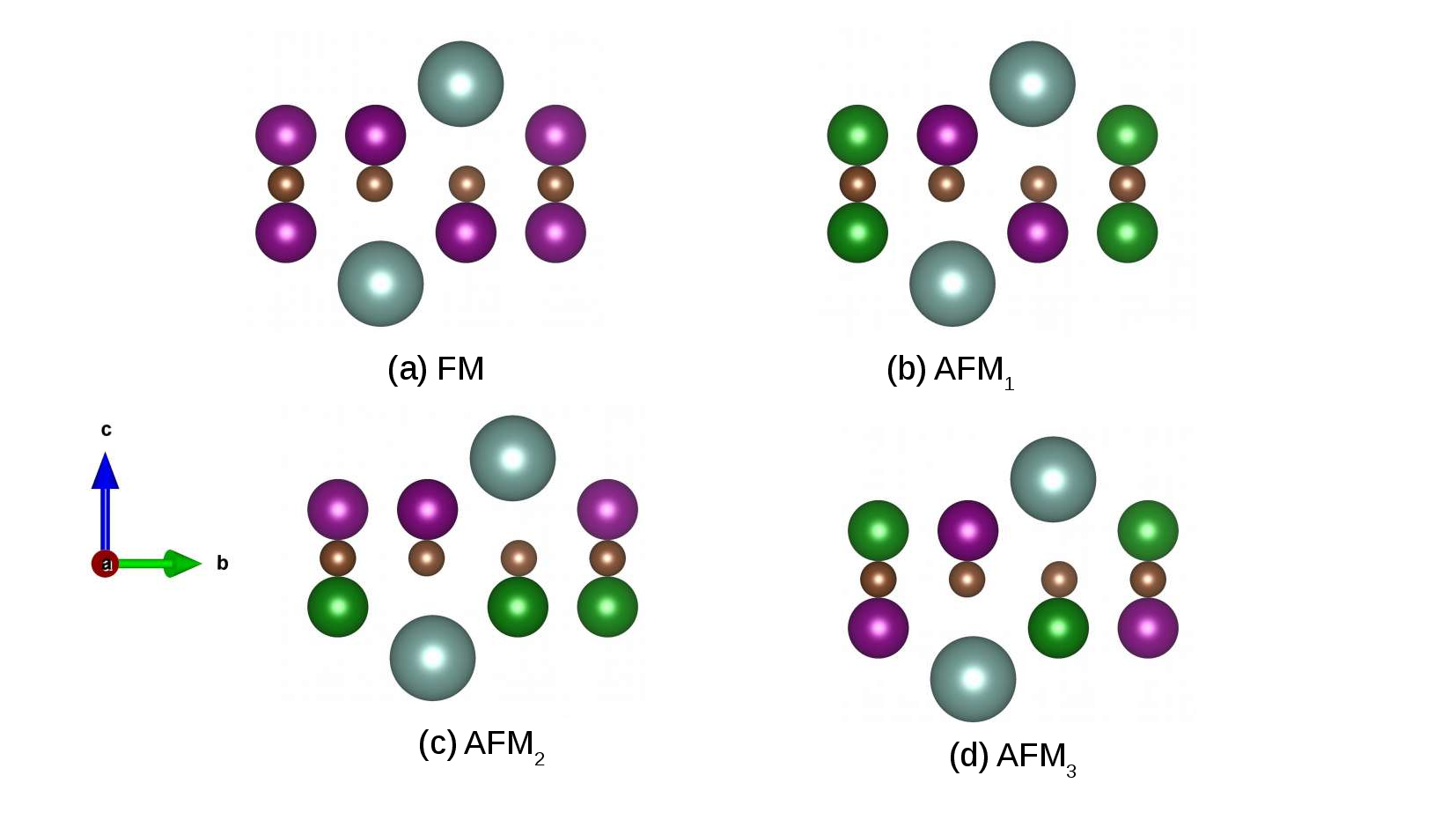}
     \caption{Four magnetic configurations considered in this work.The purple(green) ball indicates spin of the Cr atom pointing along (against) the $c$-axis. Brown and grey balls depict C and M$^{\prime}$ atoms, respectively.}. 
    \label{Fig:2}
\end{figure*}
In Figure \ref{Fig:1}, the structure of pristine i-MXene (Cr\textsubscript{2/3}M$^{\prime}$\textsubscript{1/3})\textsubscript{2}C (M$^{\prime}$=Sc/Zr/Y) is shown. Pristine (Cr\textsubscript{2/3}M$^{\prime}$\textsubscript{1/3})\textsubscript{2}C is essentially a trilayer structure, with a single layer of C atoms sandwiched between two layers of Cr/M$^{\prime}$ atoms. The M$^{\prime}$ atoms are slightly misaligned with respect to the Cr atoms (Figure \ref{Fig:1}(b)). The structure of the conventional cell of i-MXene considered in this work is  taken from Reference \onlinecite{gao2020magnetic}. In order to assess the possibility of exfoliation of the i-MXenes  from their corresponding iMAX phases, we first calculated their exfoliation energies($E_{exf}$). The calculated exfoliation energies are shown in Figure S1, supplementary information. The exfoliation energies are significantly less than that of the well-known and commercially synthesized MXene Ti\textsubscript{3}C\textsubscript{2}; the value being 0.158 eV/\AA$^{2}$… \cite{khazaei2018insights} for exfoliating from Ti\textsubscript{3}AlC\textsubscript{2}. This suggests that the  i-MXenes considered in this work can be easily exfoliated from the corresponding  iMAX phases. 
The structures of functionalised i-MXenes can be obtained by considering the total energies of various structural models of functionalisations. The functional groups passivating their surfaces have three different positions to occupy. Two of them, denoted as A and B, are shown in Figure \ref{Fig:1}(a); A(B) represents the hollow site above the C(Cr/M$^{\prime}$) atom. The third position D (not shown in the figure) is the one right above the transition metal atom site. Thus, there can be four different structural models of surface functionalisation: BB (AA) model, where functional groups occupy hollow site B(A) of both surfaces; AB model, where functional groups attach to hollow site A of one surface and hollow site B of the other surface, or vice versa; and DD model, where the functional groups occupy the D sites associated with both surfaces. Total energies of  these models for each i-MXene considered, are provided in Tables I-III, supplementary information. The results suggest that for all three compounds, irrespective of the functional groups, the lowest energy configuration is obtained in BB model. After obtaining the optimized structures of the functionalised i-MXenes, formation energy of each one of them is calculated to confirm their stability. The results are shown in Fig S2, supplementary information. The large and negative formation energies indicate formation of strong chemical bonds in these systems.
\subsection{Magnetic ground states}
In order to find the magnetic ground state structure,  spin-polarized calculations with various ordered magnetic configurations are done on the stable crystal structures of functionalised MXenes. Four different magnetic configurations, illustrated in Figure \ref{Fig:2}, are considered. Apart from the ferromagnetic (FM) configuration where all Cr spins are aligned along $c$-direction, three antiferromagnetic(AFM) configurations are considered : (a)AFM\textsubscript{1} where antiferromagnetic (ferromagnetic) alignment is considered along $b(c)$-direction (b) AFM\textsubscript{2} where antiferromagnetic (ferromagnetic) alignment is considered along $c(b)$-direction and (c) AFM\textsubscript{3} where the alignment is antiferromagnetic along both $b$ and $c$ directions.  The magnetic properties of the i-MXenes considered in this work are  provided in Table \ref{Table1}.
\begin{table*}
\caption{\label{Table1}The energy $\Delta E$  (in eV per unit cell) of the three AFM configurations with respect to the energy of the FM configurations, magnetic moment per Cr atom M\textsubscript{Cr} (in Bohr-magneton $\mu_{B}$), magnetic ordering temperature T \textsubscript{C}/T\textsubscript{N} (in K) and $E_{MAE}$ (in meV per unit cell), the magnetic anisotropy energyy are shown for pristine and functionalised i-MXenes considered in this work.}
\begin{tabular}{ m{0.19\textwidth}m{0.14\textwidth}m{0.14\textwidth}m{0.14\textwidth}m{0.14\textwidth}m{0.08\textwidth}m{0.14\textwidth}}
 \hline
 System & $ \Delta E_{AFM\textsubscript{1}-FM} $ & $ \Delta E_{AFM\textsubscript{2}-FM} $ & $ \Delta E_{AFM\textsubscript{3}-FM} $ & M\textsubscript{Cr} & T\textsubscript{C}/T\textsubscript{N} & $E_{MAE}$\\
 \hline
 (Cr\textsubscript{2/3}Y\textsubscript{1/3})\textsubscript{2}C & -39.59  & -6.88 &  -109.8  & 1.1  & 55 & 0.375 \\
 (Cr\textsubscript{2/3}Sc\textsubscript{1/3})\textsubscript{2}C & -11.5 & -9.4 & -21.63 & 1.15 & 30 & 0.344\\
 (Cr\textsubscript{2/3}Zr\textsubscript{1/3})\textsubscript{2}C & -53 & -46.81 & -46.81  & 0.96  & 30  & 0.301 \\
 (Cr\textsubscript{2/3}Y\textsubscript{1/3})\textsubscript{2}CF\textsubscript{2} & 220 & -355.30  & 82.78  & 1.92 & 85 & 0.164 \\
 (Cr\textsubscript{2/3}Sc\textsubscript{1/3})\textsubscript{2}CF\textsubscript{2} & 195.8 & -555.3 & 56.2  & 2.06 & 140 & -0.009\\
 (Cr\textsubscript{2/3}Zr\textsubscript{1/3})\textsubscript{2}CF\textsubscript{2} & 190.20 & -189.46 & 190.20  & 1.85 & 155 & 0.362 \\
 (Cr\textsubscript{2/3}Y\textsubscript{1/3})\textsubscript{2}CO\textsubscript{2} & 369.7 & 366 & 313.41  & 1.54 & 115 & 0.034\\
 (Cr\textsubscript{2/3}Sc\textsubscript{1/3})\textsubscript{2}CO\textsubscript{2} & 178.28 & 278.02 & 208.40 & 1.48 & 70 & 0.157 \\
  (Cr\textsubscript{2/3}Zr\textsubscript{1/3})\textsubscript{2}CO\textsubscript{2} & -10.36 & 11.50 & 143.32  & 1.73,0.72 & 50 & 0.076 \\
 \hline
\end{tabular}
\end{table*}
The results clearly show that surface passivation influences the magnetic ground states of these i-MXenes. The pristine  iMXenes,  (Cr\textsubscript{2/3}Sc\textsubscript{1/3})\textsubscript{2}C, (Cr\textsubscript{2/3}Y\textsubscript{1/3})\textsubscript{2}C, and (Cr\textsubscript{2/3}Zr\textsubscript{1/3})\textsubscript{2}C, stabilise in antiferromagnetic (AFM) ground states although the AFM configurations vary across systems. While (Cr\textsubscript{2/3}Y\textsubscript{1/3})\textsubscript{2}C and (Cr\textsubscript{2/3}Sc\textsubscript{1/3})\textsubscript{2}C stabilise in AFM\textsubscript{3} configuration, (Cr\textsubscript{2/3}Zr\textsubscript{1/3})\textsubscript{2}C stabilises in  AFM\textsubscript{1} state. The former two compounds have near identical Cr moments while that in (Cr\textsubscript{2/3}Zr\textsubscript{1/3})\textsubscript{2}C is slightly less. This suggests identical mechanism driving the magnetic ground states in (Cr\textsubscript{2/3}Y\textsubscript{1/3})\textsubscript{2}C and (Cr\textsubscript{2/3}Sc\textsubscript{1/3})\textsubscript{2}C. 
 Identical magnetic ground states are obtained in -F functionalised i-MXenes. -F functionalisation stabilises the AFM\textsubscript{2} order in all three systems considered. The Cr magnetic moment increases by about 1 $\mu_{B}$ as compared to that in the pristine compounds. With -O functionalisation, the ground state magnetic structures change again. (Cr\textsubscript{2/3}Y\textsubscript{1/3})\textsubscript{2}CO$_{2}$ and (Cr\textsubscript{2/3}Sc\textsubscript{1/3})\textsubscript{2}CO$_{2}$ now stabilise in FM states. (Cr\textsubscript{2/3}Zr\textsubscript{1/3})\textsubscript{2}CO$_{2}$, on the other hand,  stabilises in  AFM\textsubscript{1} state. However, in this magnetic structure, the magnetic moments of the two Cr atoms are not identical: moment of the Cr atom with spin aligned(anti-aligned) along $c$-direction is 1.73(0.72) $\mu_{B}$. Thus, the ground state of (Cr\textsubscript{2/3}Zr\textsubscript{1/3})\textsubscript{2}CO$_{2}$ is Ferrimagnetic (FiM). The Cr moments of -O functionalised i-MXenes are found to be lying in between those of pristine and the -F functionalised ones. 
 
 That the functional group passivating the surfaces along with the alloying at the transition metal site affects the magnetic order in MXenes are evident from these results. A comparison with existing results for conventional MXenes Cr\textsubscript{2}CF\textsubscript{2} \cite{si2015half} and Cr\textsubscript{2}CO\textsubscript{2} \cite{khazaei2013novel} corroborates this. While Cr\textsubscript{2}CF\textsubscript{2} stabilises in an AFM structure, Cr\textsubscript{2}CO\textsubscript{2} has a non-magnetic ground state. The emergence of significant Cr moment in -O functionalised i-MXenes, therefore, is due to alloying with a non-magnetic transitional metal atom. The difference between conventional and i-MXene in case of -F functionalised systems considered here is reflected in the considerable reduction of the Cr moments. The DFT calculated Cr moments in Reference \onlinecite{si2015half} are 3 $\mu_{B}$, 50$\%$ more than the results obtained for the i-MXenes in the present work. Even in case of pristine compounds, the ground states and the magnetic moments of (Cr\textsubscript{2/3}M$^{\prime}$\textsubscript{1/3})\textsubscript{2}C are different from conventional MXene Cr$_{2}$C that is found to be stabilised in FM configuration with a magnetic moment of 4 $\mu_{B}$ per Cr atom. The effects of functionalisation of surfaces in i-MXenes can also be observed by looking at the spin density profiles. To this end, we have plotted the spin density profiles of pristine and functionalised (Cr\textsubscript{2/3}Sc\textsubscript{1/3})\textsubscript{2}C (Figure S3, supplementary information). The spin density profiles indicate that while in pristine and -F functionalised i-MXenes, the spin density is localised on the Cr atoms, delocalisation has happened for the -O functionalised system. In this system, spin density is induced on the C atoms via double exchange mechanism. 
 
\subsection{Electronic Structure}
\begin{figure*}
    \includegraphics[height=5cm, width=16.00 cm]{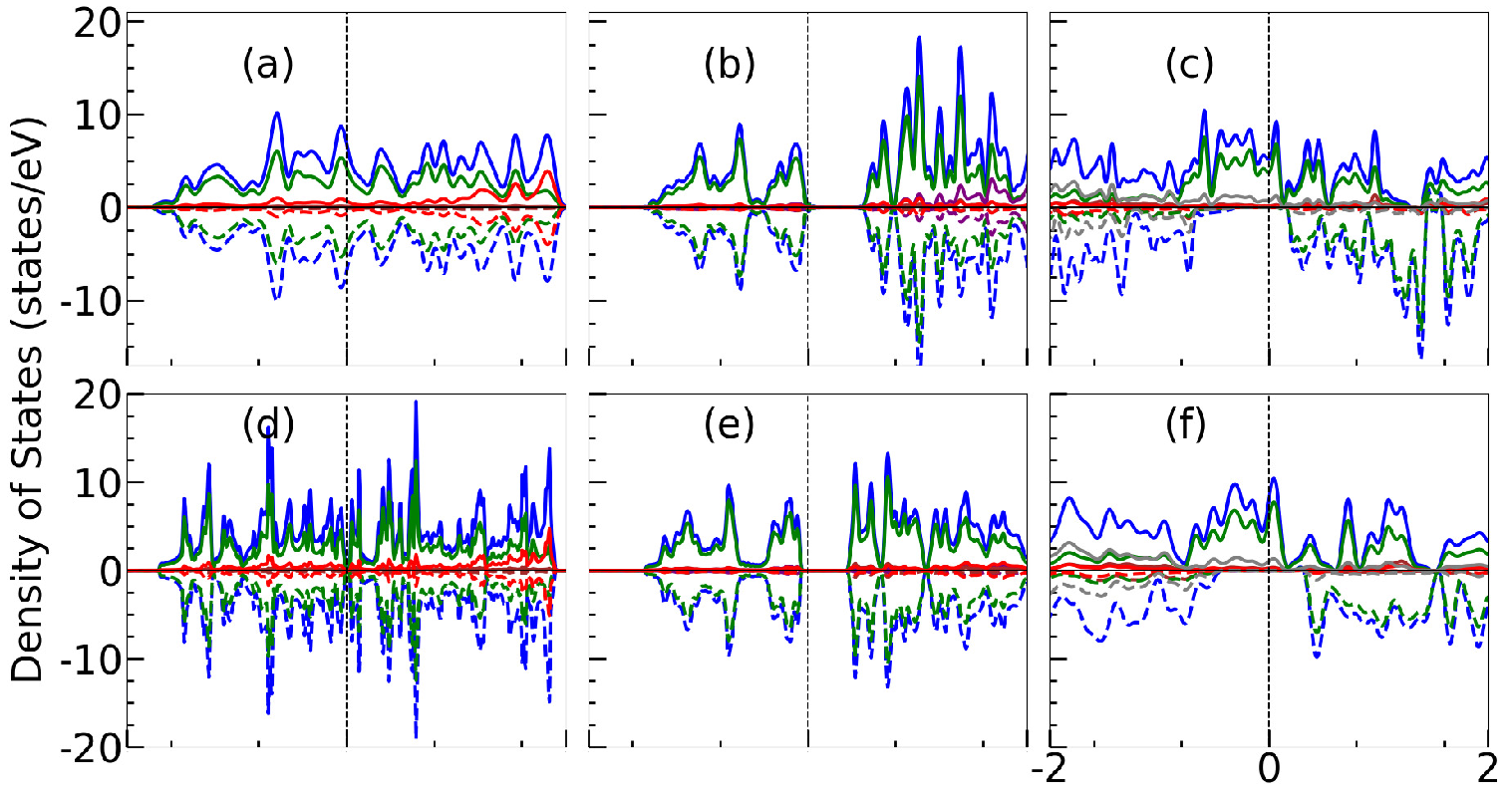}
    \vspace{0.002cm}  
    \includegraphics[height=5cm, width=16.00 cm]{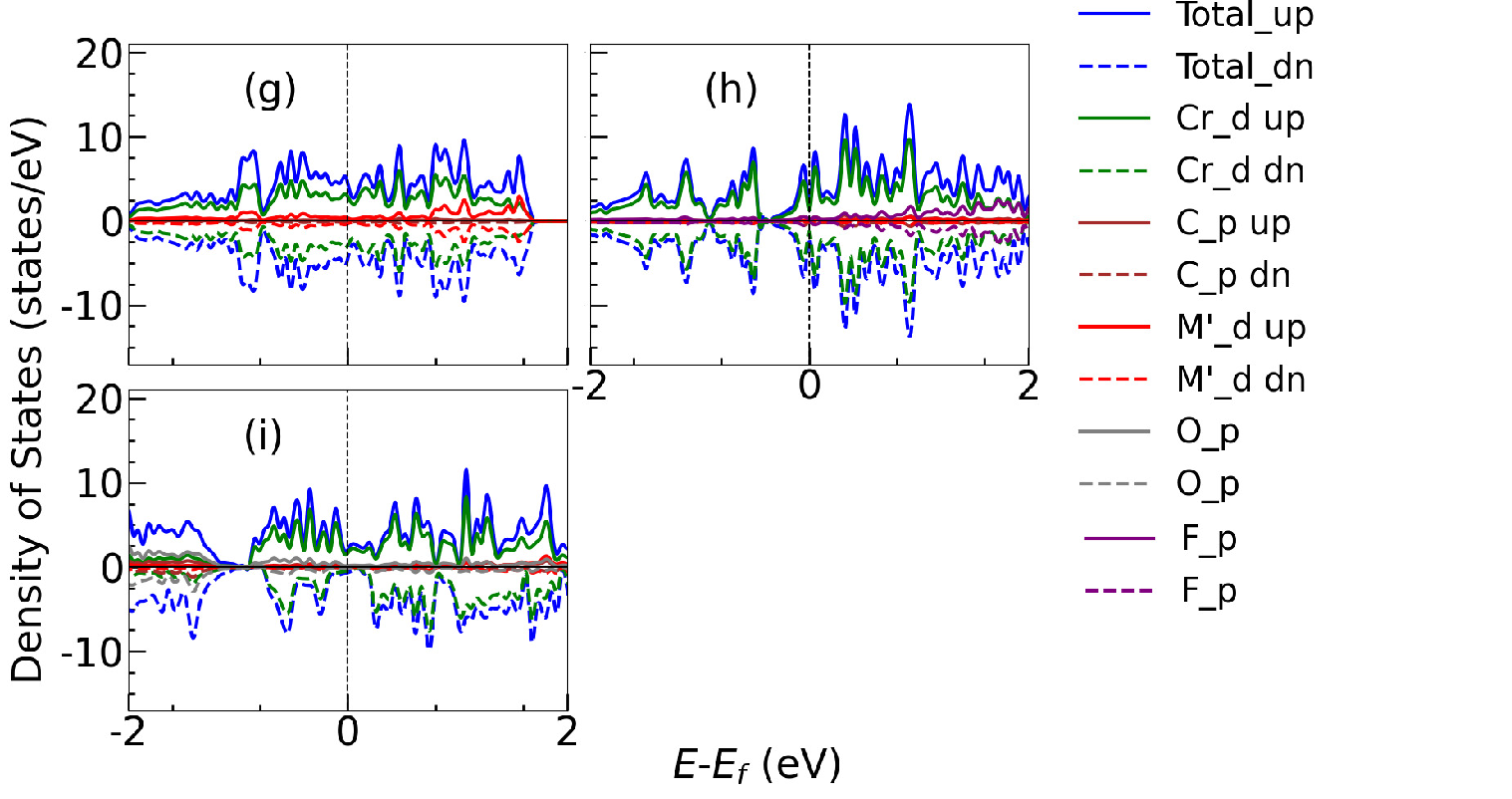} 
     \caption{Total,atom-projected and orbital-projected densities of states of (a) (Cr\textsubscript{2/3}Sc\textsubscript{1/3})\textsubscript{2}C,(b) (Cr\textsubscript{2/3}Sc\textsubscript{1/3})\textsubscript{2}CF\textsubscript{2}, (c) (Cr\textsubscript{2/3}Sc\textsubscript{1/3})\textsubscript{2}CO\textsubscript{2}, (d)(Cr\textsubscript{2/3}Y\textsubscript{1/3})\textsubscript{2}C, (e) (Cr\textsubscript{2/3}Y\textsubscript{1/3})\textsubscript{2}CF\textsubscript{2},(f) (Cr\textsubscript{2/3}Y\textsubscript{1/3})\textsubscript{2}CO\textsubscript{2},(g) (Cr\textsubscript{2/3}Zr\textsubscript{1/3})\textsubscript{2}C,(h)(Cr\textsubscript{2/3}Zr\textsubscript{1/3})\textsubscript{2}CF\textsubscript{2}  and (i) (Cr\textsubscript{2/3}Zr\textsubscript{1/3})\textsubscript{2}CO\textsubscript{2}. "up" and "down" indicate the orientations of the spins with respect to $c$-direction. Contributions from $d$ orbitals of transition metals Cr,M$^{\prime}$ and $p$ orbitals of C and the functional groups are plotted along with the total densities of states.}
    \label{Fig:3}
\end{figure*} 
\begin{figure*}
    \includegraphics[height=9cm, width=17.00 cm]{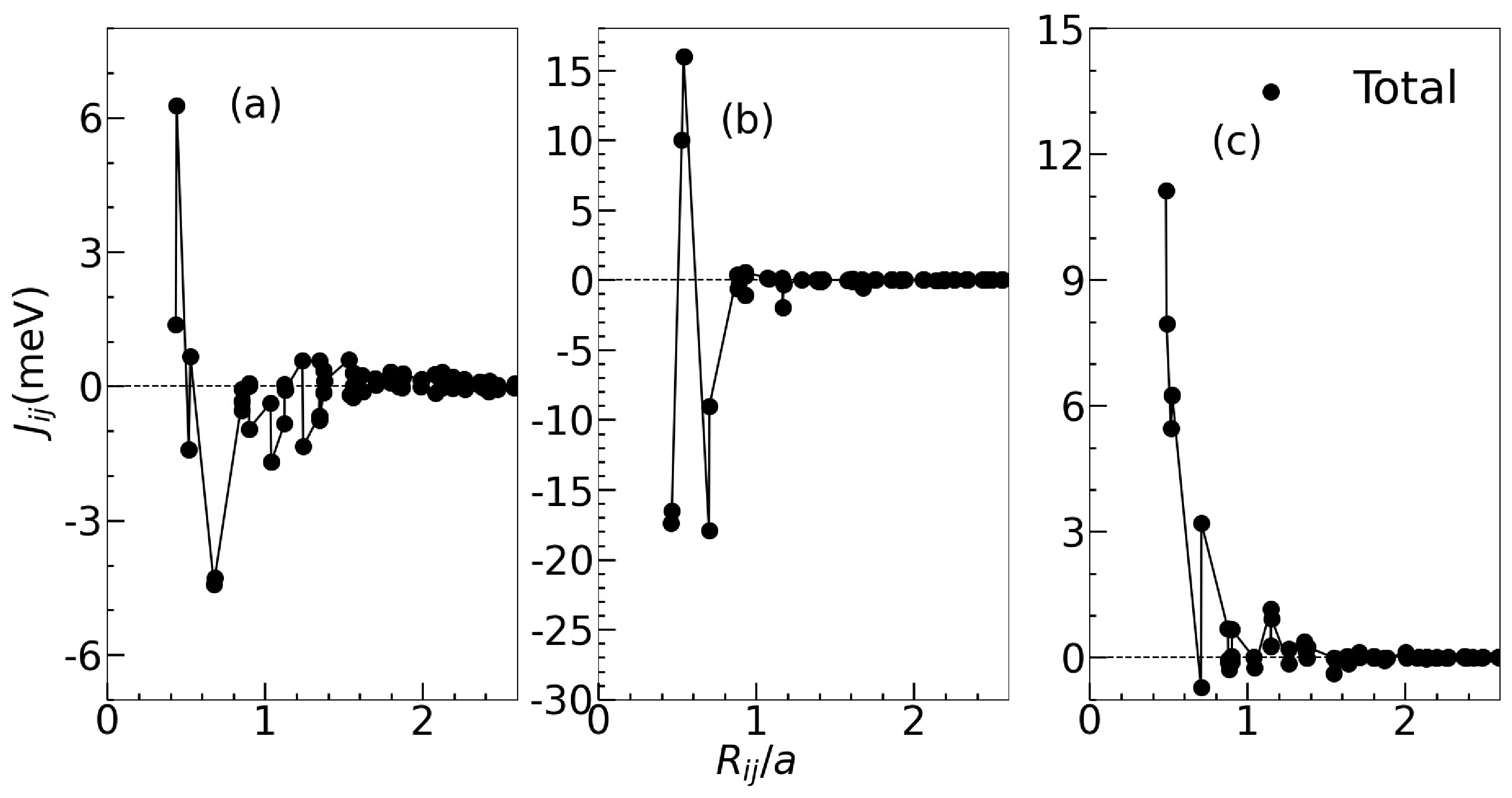}
     \caption{Inter-atomic exchange parameters as a function of inter-atomic distances for (a) (Cr\textsubscript{2/3}Sc\textsubscript{1/3})\textsubscript{2}C, (b) (Cr\textsubscript{2/3}Sc\textsubscript{1/3})\textsubscript{2}CF\textsubscript{2} and (c) (Cr\textsubscript{2/3}Sc\textsubscript{1/3})\textsubscript{2}CO\textsubscript{2}.}
    \label{Fig:4}
\end{figure*}

The total, atom and orbital-projected spin polarised densities of states of the systems considered are shown in Figure \ref{Fig:3}, and Figure S4, supplementary information. Comparison of the three pristine i-MXenes (Figure \ref{Fig:3}(a),(d) and (g),indicates that while (Cr\textsubscript{2/3}Sc\textsubscript{1/3})\textsubscript{2}C and (Cr\textsubscript{2/3}Zr\textsubscript{1/3})\textsubscript{2}C are metallic, (Cr\textsubscript{2/3}Y\textsubscript{1/3})\textsubscript{2}C is a semiconductor with a small band gap of 20 meV.The atom-projected densities of states  indicate strong hybridisation between $d$-orbitals of the transition metal constituents and $p$-orbitals of C. The differences in the electronic ground states of the three compounds can be understood from the orbital-decomposed densities of states shown in Figure S4, supplementary information.  In the pristine i-MXenes, each transition metal atom sits in  a C\textsubscript{3v}  crystal field. Under the C\textsubscript{3v} symmetry, the $d$-orbitals of transition metal atoms split into  $ a_{1} (d_{z^{2}}) $, doubly degenerate $ e_{1} (d_{xz} , d_{yz}) $ and $ e_{2} ( d_{x^{2}-y^{2}} , d_{xy}) $ orbitals. In the pristine case, the $e_{1}$ states are higher in energy than the other two. The relative positions of the three orbitals decide the electronic ground state. In case of (Cr\textsubscript{2/3}Y\textsubscript{1/3})\textsubscript{2}C, the distinction in relative positions of the three orbitals is very clear. Unlike the other two pristine compounds, the localised orbitals in this compound is responsible for the semiconducting ground state. 

 F-functionalization localises the $d$-states of transition metals resulting in semiconducting ground states in (Cr\textsubscript{2/3}Y\textsubscript{1/3})\textsubscript{2}CF\textsubscript{2} and (Cr\textsubscript{2/3}Sc\textsubscript{1/3})\textsubscript{2}CF\textsubscript{2}, the band gaps being 0.565 eV and 0.679 eV, respectively. The deep lying -F states push the $d$-states in the occupied part of the spectrum towards lower energies, opening the gaps in both spin channels. The high magnetic moments of Cr atoms in -F functionalised i-MXenes is an artefact of this. Although localisation is also observed in (Cr\textsubscript{2/3}Zr\textsubscript{1/3})\textsubscript{2}CF\textsubscript{2}, the states in the unoccupied parts too are pushed substantially to the lower energies resulting in a metallic ground state. This is primarily due to the delocalised $a_{1}$ states in the bottom of the conduction bands. It is worth mentioning that in case of -F functionalised i-MXenes, the energy hierarchy of the $d$-orbitals is different than that of the pristine cases. In the -F functionalised i-MXenes, the transition metal atoms are in an octahedral environment with D\textsubscript{3d} symmetry. Due to this the $e_{1}, e_{2}$ orbitals in the occupied part have higher energies than $a_{1}$. The reverse happens in the unoccupied part of the spectrum. 
 
All three -O functionalised i-MXenes turn out to be half-metallic. The densities of states of (Cr\textsubscript{2/3}Y\textsubscript{1/3})\textsubscript{2}CO\textsubscript{2} and (Cr\textsubscript{2/3}Sc\textsubscript{1/3})\textsubscript{2}CO\textsubscript{2} have an empty Cr spin down band. The spin down bands of M$^{\prime}$ are pushed towards lower energies opening up a semiconducting gap. This arises due to hybridisations with the deep lying O $p$ states. In case of (Cr\textsubscript{2/3}Zr\textsubscript{1/3})\textsubscript{2}CO\textsubscript{2}, the half-metallic gap is much smaller. This is due to different exchange splittings of two Cr atoms leading to more states closer to the Fermi level. The O states in all three cases are delocalised corroborating the results from the spin density profiles. This trait of -O functionalised i-MXenes considered in this work is important from the utility point of view as these materials are promising candidates as spin filters. 

The features in the electronic structures help understand the origin of the magnetic ground states of the three i-MXenes, obtained in our calculations. The localised nature of -F states in -F functionalised MXenes indicate that the magnetic interactions among the Cr atoms is predominantly superexchange stabilising an AFM configuration. On the other hand the delocalised -O states in -O functionalised i-MXenes give rise to a double exchange mechanism stabilising the ferromagnetic ground states in two of the systems considered. In the pristine iMXene, magnetism is determined by the interplay of direct exchange and superexchange interactions between Cr atoms. Due to the dominating direct exchange interactions between Cr-atoms, pristine i-MXenes has antiferromagnetic interaction along b-axis.  So the pristine iMXenes have AFM\textsubscript{1} or AFM\textsubscript{3} as the ground state .
\subsection{Interatomic exchange parameters and Transition temperatures}

In Figures \ref{Fig:4} and S5, supplementary information, we show the Cr-Cr magnetic exchange parameters as a function of inter-atomic distances. Results reveal that only the first few nearest neighbors make significant contributions to the total interactions, while long-range interactions ($ R_{ij}/a > 2.5 $) are negligibly small. Figure \ref{Fig:4}(a) shows the exchange parameters for (Cr\textsubscript{2/3}Sc\textsubscript{1/3})\textsubscript{2}C. The figure suggests that there are significant AFM interactions from the third, fifth and sixth nearest neighbours that counter the FM interactions in the first two nearest neighbour shells leading to an AFM ground state in this compound. In the -F functionalised compound (Cr\textsubscript{2/3}Sc\textsubscript{1/3})\textsubscript{2}CF\textsubscript{2} (Figure \ref{Fig:4}(b)), the magnetic interactions among first and second nearest neighbour Cr atoms are strongly AFM as is the case with the  fifth nearest neighbours. These collectively overcome strong FM interactions between the third and fourth nearest neighbour Cr atoms and stabilise an AFM ground state. Strong FM interactions is found among the first four neighbours of a Cr atom in (Cr\textsubscript{2/3}Sc\textsubscript{1/3})\textsubscript{2}CO\textsubscript{2} (Figure \ref{Fig:4}(c)), explaining the origin of an FM ground state in this compound. Same trends are found in case of the other two i-MXenes with the exception of (Cr\textsubscript{2/3}Zr\textsubscript{1/3})\textsubscript{2}CO$_{2}$ (Figure S5(f), supplementary information) where  a few interactions are AFM. However, this is consistent with the FiM ground state of this compound. 

\begin{figure*}
    \includegraphics[height=12 cm, width=18.00 cm]{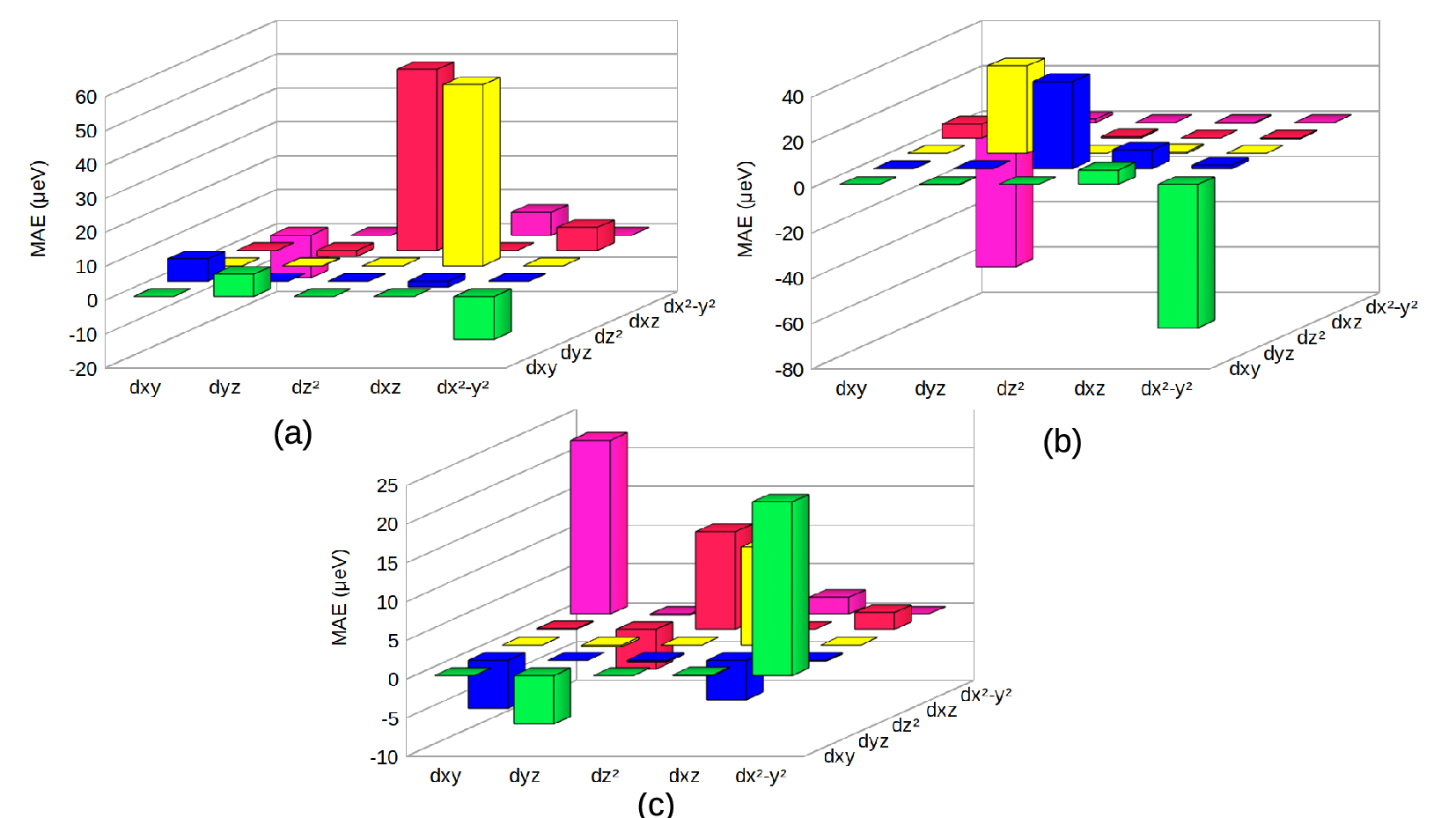}
     \caption{Orbital resolved MAE for (a) (Cr\textsubscript{2/3}Sc\textsubscript{1/3})\textsubscript{2}C, (b) (Cr\textsubscript{2/3}Sc\textsubscript{1/3})\textsubscript{2}CF\textsubscript{2} and (c) (Cr\textsubscript{2/3}Sc\textsubscript{1/3})\textsubscript{2}CO\textsubscript{2}.}
    \label{Fig:5}
\end{figure*}
These exchange interactions are used to calculate the magnetic transition temperatures employing Monte Carlo simulation. The calculated values are given in Table \ref{Table1}. We find that the transition temperatures are higher for functionalised i-MXenes in comparison with the pristine compounds; the highest transition temperatures are achieved in case of -F functionalised i-MXenes. This is due to strongest superexchange between the Cr atoms in these compounds. The transition temperatures of these i-MXenes are comparable to few of the established 2D magnetic materials like CrI$_{3}$, FePS$_{3}$ and Cr$_{2}$Ge$_{2}$Te$_{6}$ (Figure S6, supplementary material). 

\subsection{Magnetic anisotropy energy }
Magnetic anisotropy energy (MAE) of a material is important from the point of view of device application as it is instrumental in stabilisation of  the magnetic long-range order in the system. The  MAE is calculated the following way:\\
  $$ E_{MAE} = E_{||}- E_{\perp} $$ 
$E_{||}$ and $E_{\perp}$ represent the total energy of the magnetic system when the spins are aligned in-plane and out-of-plane with respect to the surface, respectively. Spin orbit coupling is included in the calculations to obtain MAE from DFT total energies. The in-plane energies are calculated along the (100) and (010) directions while the out-of-plane energies are calculated along (001) direction. Results for the i-MXenes considered in this work are presented in Table \ref{Table1}. A positive MAE value signifies the presence of an easy axis perpendicular to the MXene surface, while a negative sign indicates that the easy direction aligns parallel to the MXene surface.
Our results imply that for pristine MXenes, the MAE values are near identical, large; the magnetisation easy axis for all of them being along (001). 
Among the F-functionalised iMXenes, the magnitudes of MAE differ significantly. While (Cr\textsubscript{2/3}Zr\textsubscript{1/3})\textsubscript{2}CF\textsubscript{2} has MAE comparable to the pristine compounds, the other two have noticeable lower values. Moreover, result for  (Cr\textsubscript{2/3}Sc\textsubscript{1/3})\textsubscript{2}CF\textsubscript{2} implies an in-plane magnetisation. O-functionalisation reduces MAE values further. However, all three O-functionalised compounds have (001) as magnetisation easy axis. In Figure S7, we compare our calculated MAE values for the three i-MXenes with those of a few established 2D magnetic materials. We find that though the calculated MAE values of these i-MXenes are not as large as those obtained for CrI$_{3}$ or VSe$_{2}$, they are comparable with or larger than the MAE of other investigated 2D magnets like CrBr$_{3}$ and Cr$_{2}$Te$_{2}$Ge$_{6}$. 

 The MAE of the i-MXenes considered, predominantly come from the Cr atoms. The contributions from other atoms are negligible. This is due to the fact that the  states near the Fermi level majorly contribute to the MAE. This, also, is the reason for MAE getting substantially modified upon functionalisation of the pristine i-MXenes. Thus,to get an insight into the origin of MAE in these i-MXenes, investigations into the contributions of various Cr $d$ orbitals towards their MAE can be useful.
 The MAE associated with a transition metal can be understood based on spin orbit coupling (SOC) interactions between the occupied and unoccupied d-states. By treating this interaction as perturbation through second order perturbatiion theory this MAE can be estimated as  \cite{lee2022out,wang1993first}
\begin{align*}
\resizebox{\hsize}{!}{$E_{MAE} = \xi^{2}\sum_{o,u,\alpha,\beta} (-1)^{1-\delta_{\alpha \beta}}\frac{\mid \langle o^{\alpha}\mid L_{z}\mid u^{\beta}\rangle\mid^{2} - \mid \langle o^{\alpha}\mid L_{x}\mid u^{\beta}\rangle\mid^{2}}{\epsilon^{\beta}_{u} - \epsilon^{\alpha}_{o}}$}.
\end{align*} 
$\xi $ is SOC constant, $ L_{x}(L_{z}) $ is the angular momentum operator,$ u (o)$ denotes the unoccupied (occupied)states, $ \alpha $ and $\beta $ denote spin component  and $ \epsilon $ is the eigenenergy, Denoting the SOC interactions between the Cr $3d$ orbitals as $ d_{i}\otimes d_{j}$ \cite{lee2022out}, contributions of various orbitals $i$ towards MAE are illustrated in Figure \ref{Fig:5} and S8, supplementary information.

For the pristine compounds, large positive contribution comes from $ d_{z^2}\otimes d_{xz} $ interactions. For (Cr\textsubscript{2/3}Y\textsubscript{1/3})\textsubscript{2}C, additional positive contribution comes from $d_{xz} \otimes d_{x^{2}-y^{2}}$ interactions (Figure S8(a), supplementary information). This additional interaction produces the highest MAE in this compound. Interactions between the other orbitals do not have any significant contribution. Since the contributions of the dominant $ d_{z^2}\otimes d_{xz} $ interaction are nearly same, these three compounds have near identical MAE values. 
Very low and negative MAE value in (Cr\textsubscript{2/3}Sc\textsubscript{1/3})\textsubscript{2}CF\textsubscript{2} (Figure \ref{Fig:5}(b)) is due to a competition between two sets of orbital interactions: a negative contribution coming from  $ d_{xy}\otimes d_{x^{2}-y^{2}} $ and a near equal positive contribution coming from the $ d_{z^2}\otimes d_{yz} $ interactions. Similarly, relatively smaller but positive MAE found in (Cr\textsubscript{2/3}Sc\textsubscript{1/3})\textsubscript{2}CO\textsubscript{2} is due to relatively smaller positive contributions from $ d_{xy}\otimes d_{x^2 - y^2} $ and $ d_{z^2}\otimes d_{xz}$ interactions and non-negligible negative contributions from $ d_{xy}\otimes d_{yz} $ and $ d_{yz}\otimes d_{xz} $ interactions (Figure \ref{Fig:5}(c)). The decrease in MAE values for -O functionalised (Cr\textsubscript{2/3}Y\textsubscript{1/3})\textsubscript{2}C and (Cr\textsubscript{2/3}Zr\textsubscript{1/3})\textsubscript{2}C by one order of magnitude as compared to (Cr\textsubscript{2/3}Sc\textsubscript{1/3})\textsubscript{2}CO\textsubscript{2} is due to smaller(larger) positive(negative) contributions from $d_{xy}\otimes d_{x^2 - y^2} $($  d_{z^2}\otimes d_{xz} $) interactions. MAE values of the remaining two -F functionalised i-MXenes can be understood in a similar way.

\section{Conclusions}
The structural and composition flexibility of 2D MXenes is exploited in this work to explore possibility of obtaining tunable magnetism in Cr-based MXenes. Using Density functional theory based calculations, we have investigated the electronic and magnetic properties of three Cr based MXenes, known as i-MXenes, which order in-plane with the transition metal site having an alloy of Cr and another early transition metal non-magnetic constituent with a fixed composition ratio. Our calculations demonstrate that the electronic and magnetic properties of these compounds are significantly influenced by the  surface functionalisation, the chemical nature of the  functional group and the non-magnetic constituent. Thus the electronic and magnetic properties can be tuned by changing the non-magnetic transition metal component and the functional group. For the three compounds considered in this work, the -O functionalistion stabilises Ferromagnetic half-metallic states. This makes these compounds suitable as spin-filters. By analysing the electronic structures and the magnetic exchange parameters we explain the reasons behind occurrence of different magnetic ordering upon changes in the compositions. Our results on magnetic transition temperatures and magnetic anisotropy energies show that the quantities in these systems are comparable to quite a few established 2D magnets. This implies that these i-MXenes can be useful as magnetic devices. Specifically,  (Cr\textsubscript{2/3}Zr\textsubscript{1/3})\textsubscript{2}CF\textsubscript{2} with magnetic anisotropy energy comparable to MnSe$_{2}$ and magnetic transition temperature comparable to Cr$_{2}$S$_{3}$ and CrSe can have multiple utilities as device. With increasing exploration of MXenes for magnetic applications, this work, the first one to investigate the structure-property relationships in functionalised i-MXenes, can serve as useful input to the experimentalists.  

\section{Acknowledgement}
The authors gratefully acknowledge the Department of Science and Technology, India, for the computational facilities under Grant No. SR/FST/P-II/020/2009 and IIT Guwahati
for the PARAM supercomputing facility.
\bibliographystyle{rsc} 

\providecommand*{\mcitethebibliography}{\thebibliography}
\csname @ifundefined\endcsname{endmcitethebibliography}
{\let\endmcitethebibliography\endthebibliography}{}

\end{document}